\documentclass[aps, a4paper,superscriptaddress, nofootinbib, twocolumn]{revtex4}

\usepackage{amsmath,amssymb,bm,natbib}
\usepackage{epsfig}
\usepackage{graphicx}
\usepackage{amsmath}
\usepackage{color}

\newcommand{\be}{\begin{equation}}
\newcommand{\ee}{\end{equation}}
\newcommand{\beq}{\begin{equation}}
\newcommand{\eeq}{\end{equation}}
\newcommand{\bea}{\begin{eqnarray}}
\newcommand{\eea}{\end{eqnarray}}

\newcommand{\D}{\displaystyle}
\newcommand{\gev}{\, \text{GeV}}
\newcommand{\mev}{\, \text{MeV}}

\newcommand{\disc}{{\rm disc}\,}

\renewcommand{\Im}{{\rm Im}\,}

\begin{document}

\title{Testing the consistency of the \boldmath{$\omega\pi$} transition form factor with unitarity and analyticity}

\author{I. Caprini} 
\affiliation{Horia Hulubei National Institute for Physics and Nuclear Engineering,
   P.O.B. MG-6, 077125 Bucharest-Magurele, Romania}

\begin{abstract}{
We perform a  dispersive analysis of the  $\omega\pi$ electromagnetic transition form factor, using as input the discontinuity provided by unitarity below the $\omega\pi$ threshold and including for the first time  experimental data on the modulus measured from $e^+e^-\to\omega\pi^0$  at higher energies. The input leads to  stringent  parameterization-free  constraints on the modulus of  the form factor below the $\omega\pi$  threshold, which are in  disagreement with some experimental values  measured from $\omega\to \pi^0\gamma^*$ decay. We discuss the dependence on the input parameters in the unitarity relation, using for illustration an $N/D$ formalism for the P partial wave of the scattering process $\omega\pi \to \pi\pi$, improved by a simple prescription which simulates the  rescattering in the crossed channels. Our results confirm the existence of a conflict between experimental data and theoretical calculations of the  $\omega\pi$  form factor in the region around $0.6\gev$ and bring further arguments in support of renewed experimental efforts to measure more precisely the $\omega\to\pi^0\gamma^*$ decay. }
\end{abstract}

\maketitle
\section{Introduction}

The transition form factors of light mesons play an important role in low energy precision tests of QCD \cite{Czerwinski:2012ry}. In particular, they enter as contributions to  hadronic light-by-light scattering 
calculations \cite{roadmap}, which are crucial for a more accurate theoretical determination of the standard model prediction for the muon's anomalous magnetic moment (for recent reviews see \cite{Jegerlehner:2009ry}, \cite{Benayoun:2014tra}).  

 The case of the $\omega\pi$ electromagnetic form factor is particularly interesting as there are  some discrepancies between the theoretical calculations and the experimental data from the decay $\omega\to\pi^0\gamma^*$ reported in \cite{LeptonG,NA60,NA60new}. This form factor was described by Vector Meson Dominance (VMD) model and by a chiral Lagrangian approach in \cite{Terschluesen:2010ik,Terschlusen:2012xw}. Calculations based on a standard dispersion relation were performed a long time ago
 in \cite{Koepp:1974da} and recently  in \cite{Schneider:2012ez, Danilkin:2014cra}. The discontinuity of the form factor required in the Cauchy integral can be expressed in terms of known observables by using unitarity.  The two-pion contribution to the unitarity sum gives the discontinuity in terms of the P partial wave of the amplitude of the process $\omega\pi\to\pi\pi$, itself calculated  in the dispersion theory \cite{Koepp:1974da, Niecknig:2012sj,Danilkin:2014cra},   and the pion electromagnetic form factor, a quantity which is known with very good precision.  However, the two-pion approximation is valid only in a region which  extends to a good approximation up to the $\omega\pi$ threshold, $t_+ = (m_\omega+m_\pi)^2$.  Due to the lack of information on the discontinuity above this threshold, various assumptions  were adopted for the evaluation of  the dispersion integral, either by applying
the two-pion approximation also at higher energies \cite{Koepp:1974da, Schneider:2012ez}, or by expanding the dispersion integral in powers of a suitable variable \cite{Danilkin:2014cra}. 

 A study performed recently  in \cite{Anant:2014} used as input above the $\omega\pi$ threshold, instead of the discontinuity, a model-independent integral condition on the modulus squared of the $\omega\pi$  form factor. The condition was obtained  by using an approach  proposed originally  by Okubo \cite{Okubo}, which has come to be known as the method of unitarity
bounds (a recent review of this approach is presented in ~\cite{Abbas:2010EPJA}). It exploits unitarity and the positivity of the spectral function of a suitable current--current correlator, calculated by operator product expansion (OPE) in the Euclidean region.   In the particular case of the $\omega\pi$ form factor, the method, adapted to the specific input conditions available, led eventually to a functional optimization problem of a type considered  for the first time in \cite{Caprini:1981,Caprini:1982nr}. The solution of the problem yields upper and lower bounds on the modulus of the  $\omega\pi$  form factor in the region  $4 m_\pi^2\leq t< t_+$ \cite{Anant:2014}.  A specific feature of this form factor is that its discontinuity across the cut is not purely imaginary. As a consequence,  the form factor is not a real analytic function, as happens in familiar cases like the pion vector form factor.  Therefore, in \cite{Anant:2014} the formalism of bounds was extended to functions which are not real analytic.
Although not very stringent, the bounds derived  in \cite{Anant:2014} are in disagreement with the experimental data on the modulus of the form factor in the region around $0.6\gev$, measured from the decay $\omega\to\pi^0\gamma^*$, confirming thus the conclusion of the analysis \cite{Schneider:2012ez} based on standard dispersion theory. 

It is important to note that the dispersive analyses performed so far did not include experimental data on the form factor available in the scattering region, above the $\omega\pi$ threshold. Measurements of the modulus from the reaction $e^+e^-\to\omega\pi^0$ are reported in 
\cite{Dolinsky:1986kj,Bisello:1990du, SND-omegapi,CMD2-omegapi,Achasov:2012zz, Edwards:1999fj, KLOE:omegapi0gamma} (a set of such data is shown in Fig. \ref{fig:exp}, where we show for completeness also the modulus measured in the decay region $t<t_-$,  $t_- = (m_\omega-m_\pi)^2$).  In the present paper we consider the problem of including this information in the dispersive formalism. Specifically, we perform  an analysis of the form factor using as input the discontinuity for $t< t_+$, calculated in a theoretical model based on unitarity, and experimental information on the modulus for $t>t_+$. Even though the modulus is not known at all energies, we can implement the information in a conservative way, as a condition on a weighted integral of the modulus squared from $t_+$ to infinity.  This leads to a mathematical problem  similar to that encountered and solved in \cite{Anant:2014}. The result is expressed in the form of explicit upper and lower bounds on the modulus of the $\omega\pi$  form factor below $t_+$, calculable in terms of the discontinuity  below $t_+$ and the modulus above $t_+$. The formalism provides therefore a consistency test for the experimental data on the $\omega\pi$ electromagnetic form factor, which exploits analyticity and unitarity in a parametrization-free way.

 The theoretical  input of the test consists from the unitarity relation giving the discontinuity  of the form factor, which involves the amplitude of the $\pi\omega\to\pi\pi$ scattering. It is of interest to study  the influence of the possible uncertainties in this part of the input. The full calculation involves the solution of integral equations known as Khuri-Treiman equations \cite{KhuriTreiman}.  The amplitudes obtained in this formalism in \cite{Niecknig:2012sj,Danilkin:2014cra} are available only in numerical form. The older treatment performed in  \cite{Koepp:1974da}, based on $N/D$ formalism,  has the advantage of providing explicit expressions. However, it is not entirely satisfactory, as it does not account for the rescatterings between all the final pions in the kinematical region where the decay 
$\omega\to\pi\pi\pi$ is allowed. It is worthwhile trying
to cure the shortcomings of this approach. In this paper we consider an improved $N/D$ treatment, obtained by applying to \cite{Koepp:1974da} a prescription proposed in \cite{Moussallam:2013} for including finite-width effects in the resonances exchanged in the crossed channels. The improved model captures the essential features of the full solution, still preserving  the explicit dependence on the input parameters.   This has enabled us to investigate the influence of the uncertainties of the theoretical part of the consistency test.

The paper is organized as follows. In the next section we review the basic definitions and show how the input information on the form factor can be expressed as an extremal problem for analytic functions. 
In section \ref{sec:sol} we review the main steps of the proof and write down the solution of the extremal problem  obtained in Ref. \cite{Anant:2014}.  The results are presented in Sec. \ref{sec:res}, where we  discuss also their dependence on the parameters of the input. Sec. \ref{sec:disc} contains our conclusions. The paper has an Appendix where we describe  briefly the  $N/D$ model of \cite{Koepp:1974da} and an improved version based on a prescription suggested in  \cite{Moussallam:2013}.

\section{Input in the consistency test}\label{sec:input}
We begin with a brief description of the form factor
and the constraints that it satisfies. We use the conventions of \cite{Schneider:2012ez}, where the form factor $f_{\omega\pi}(t)$ is defined from the matrix element
\begin{equation}\label{eq:fdef}
\langle \omega(p_a,\lambda)\pi^0(p_b)| j_\mu(0) | 0 \rangle = i
\epsilon_{\mu\tau\rho\sigma}\epsilon^{\tau *}(p_a, \lambda) p_b^\rho q^\sigma f_{\omega\pi}(t),
\end{equation}
where $j_\mu$ is the isovector part of the electromagnetic current, $\lambda$ denotes the $\omega$ polarization,  $q=p_a+p_b$ and $t=q^2$.
In the convention adopted here\footnote{The dimensionless form factor $F_{\pi\omega\gamma}(t)$ defined   in \cite{Koepp:1974da} is related to the definition adopted here by $F_{\pi\omega\gamma}(t)=m_\omega f_{\omega\pi}(t) $. The form factor $F_{\pi\omega}(t)$ defined in \cite{Moussallam:2013} is dimensionless, normalized as $F_{\pi\omega}(0)=1$ and is related to our definition by $f_{\omega\pi}(t)=2\sqrt{2 \tilde C_\omega} F_{\pi\omega}(t)$, where $\tilde{C}_\omega$ is defined in Eq. (35) of \cite{Moussallam:2013}  in terms of the total width  $\Gamma_{\omega\to \pi^0\gamma}$.} the form factor $f_{\omega\pi}(t)$  has dimension of ${\rm GeV}^{-1}$.

\begin{figure}
\centering
\includegraphics[width=\linewidth, width=8.4cm]{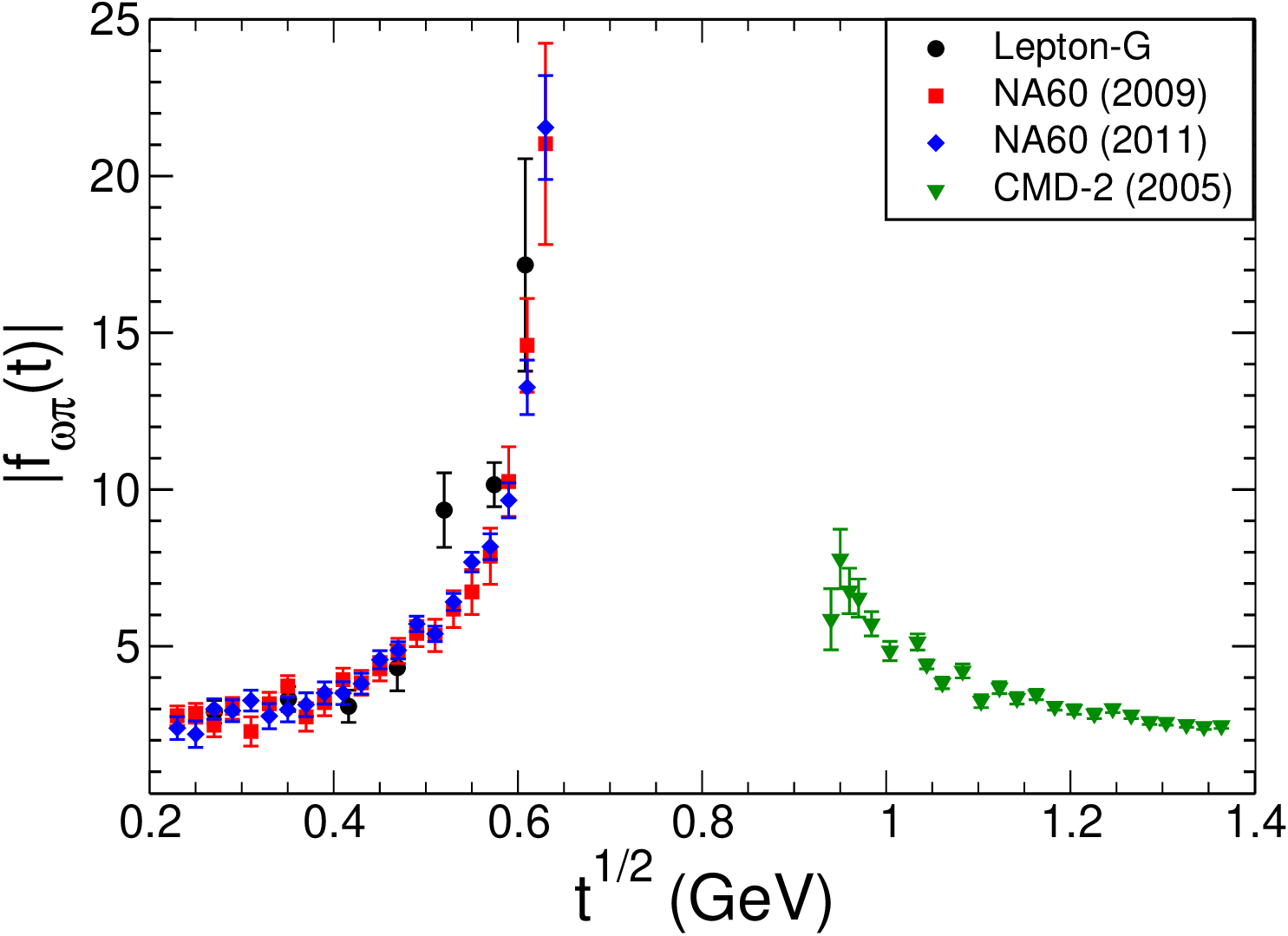}
\caption{Modulus of the $\omega\pi$ form factor measured from $\omega\to\pi^0\gamma^*$ decay  by Lepton-G~\cite{LeptonG}, NA60~(2009)~\cite{NA60} and NA60~(2011)~\cite{NA60new}, and from $e^+e^-\to \omega\pi^0 $ by CMD-2 (2005) \cite{CMD2-omegapi}.  
\label{fig:exp}}
\end{figure}

Unitarity implies that $f_{\omega\pi}(t)$ has a cut along the real axis for $t \ge 4 m_\pi^2$.
Keeping the two-pion contribution in the unitarity sum, the discontinuity of $f_{\omega\pi}(t)$ across the cut is written as
\begin{equation}\label{eq:disc}
 \disc{f_{\omega\pi}(t)}=\frac{i q(t)^3}{6\pi \sqrt{t}}F_\pi^{*}(t)f_1(t)\,\theta\big(t-4 m_\pi^2\big),\, \,\, t\leq t_+,
\end{equation}
where  $q(t)=\sqrt{t/4 - m_\pi^2}$ is the center of mass momentum of the pion pair,  $F_\pi(t)$ is the pion electromagnetic form factor and $f_1(t)$ the P~partial-wave amplitude of the scattering 
\beq\label{eq:scatt}
\omega(p_a,\lambda) \, \pi^0(p_b)\to \pi^+(q_1) \, \pi^-(q_2).
\eeq
The scattering process is physical for $t\ge t_+$.  In the region 
$4 m_\pi^2<t< t_-$, where the decay $\omega\to \pi^+\pi^-\pi^0$ is allowed,  $f_1(t)$ is the P-wave projection of the decay amplitude, while the region $t_-<t< t_+$ is unphysical. 

 The  amplitude  $f_1(t)$ was calculated in  \cite{Koepp:1974da} in the frame of $N/D$ formalism, with the left-hand cut  described by poles  in the crossed channels due to the exchange of the $\rho$ meson assumed to be stable. In this model, the phase of   $f_1(t)$ coincides with the $\pi\pi$ P-wave phase shift and exactly compensates in the discontinuity \eqref{eq:disc} the phase of $F^*_\pi(t)$, related also to the $\pi\pi$ P wave phase shift $\delta_1^1$ by Watson theorem \cite{Watson}. Therefore, the form factor $f_{\omega\pi}(t)$ calculated in \cite{Koepp:1974da} is a real analytic function\footnote{A function $F(t)$ analytic in the  $t$-plane cut for $t\ge 4 m_\pi^2$ is of real type if it satisfies the condition $F(t^*)= (F(t))^*$. In particular, this implies that the function is real on the real axis for $t<4 m_\pi^2 $, and its discontinuity across the cut can be written as $\disc F(t)\equiv  F(t+i\epsilon)- F(t-i\epsilon)=2 i \, \Im F(t+i\epsilon)$. }. 

In the more complete calculation 
\cite{Niecknig:2012sj,Schneider:2012ez,Danilkin:2014cra} based on  Khuri-Treiman formalism
\cite{KhuriTreiman}, the amplitude $f_1(t)$ is obtained by  numerically solving a set of integral equations.
Due to the rescattering between the final pions in the decay region $4 m_\pi^2<t <t_-$, the phase of $f_1(t)$ does not coincide with the $\pi\pi$ P-wave phase shift, as one would naively expect from Watson theorem. Therefore, the phases of  the two factors in \eqref{eq:disc} do not compensate each other, and the discontinuity \eqref{eq:disc} is not purely imaginary  
\cite{Schneider:2012ez,Danilkin:2014cra}. In other words, the $\omega\pi$ form factor is not a real analytic function, which is true also in the case of other transition form factors  \cite{Moussallam:2013}.

The expression \eqref{eq:disc} is valid only in the region $4 m_\pi^2\leq t < t_+$, since above the $\omega\pi$ threshold other intermediate states, besides the two-pion states, contribute in the unitarity sum. So, strictly speaking the discontinuity of $f_{\omega\pi}(t)$ is not available for $t>t_+$. On the other hand,  the modulus of $f_{\omega\pi}(t)$ can be extracted from experimental data on $e^+e^-\to\omega\pi^0$ process, measured in \cite{Dolinsky:1986kj,Bisello:1990du, SND-omegapi,CMD2-omegapi,Achasov:2012zz, Edwards:1999fj, KLOE:omegapi0gamma}. The connection between the cross section and the modulus \cite{Koepp:1974da, Moussallam:2013} is, in our convention,
\beq\label{eq:sigmaf}
\sigma_{e^+ e^-\to \omega\pi^0}(t)= \frac{ 4\pi\alpha^2}{3}\, \frac{p(t)^3 }{t^{3/2} }\, \vert
f_{\omega\pi}(t)\vert^2\,, 
\eeq
where $p(t)=\sqrt{(t-t_-)(t-t_+)/4 t}$ is the center of mass momentum of the $\omega\pi$ pair in the rest system of the virtual photon and we recall that $t_\pm= (m_\omega\pm m_\pi)^2$.  
   
Using the experimental data on the modulus and the asymptotic behaviour $|f_{\omega\pi}(t)|\sim 1/t$ predicted by perturbative QCD scaling \cite{pQCD}, it is possible to obtain a reasonable estimate  of a weighted integral over the modulus squared from $t_+$ to infinity. Thus, we can write an $L^2$-norm condition of the form
\beq\label{eq:L2}
 \frac{1}{\pi} \int_{t_+}^\infty |f_{\omega\pi}(t)|^2 \,w(t)\, dt = I,
\eeq
where $w(t)$ is a suitable weight, chosen such as to allow a precise evaluation of the quantity $I$.
We mention that a similar way of including experimental information on the modulus at higher energies was adopted in  recent investigations \cite{Ananthanarayan:2013dpa, Ananthanarayan:2013zua} of the pion electromagnetic form factor.

 In the present analysis we have considered weights of the simple form
\beq\label{eq:w}
w(t)=\frac{1}{t^c}, 
\eeq
 where the value of $c>0$ is taken such as to suppress the high-energy tail of the integral, where the form factor is not known. In practice we evaluated the quantity $I$  using an interpolation of the data on modulus from  \cite{CMD2-omegapi} shown in Fig. \ref{fig:exp} from $t_+=0.84 \,\gev^2$ up to $t=1.86\,\gev^2$, continued in a smooth way with a modulus $|f_{\omega\pi}(t)|$ decreasing like $1/t$.  As will be clear in the next section, for a fixed weight the results of the formalism   depend  monotonically on the numerical value of $I$, in the sense that a larger $I$ gives  weaker results.  Therefore, for a conservative estimate,  we have used as input in the data region the central values from  \cite{CMD2-omegapi} enlarged by their quoted errors. For $c=2$ this leads to
\beq\label{eq:I}
I =4.63\, \gev^{-4}, 
\eeq
where the region above  $1.86\,\gev^2$ contributes to the integral with about 8 \%.

Finally, we use as input the value of $|f_{\omega\pi}(0)|$, known experimentally from the $\omega\to\pi^0\gamma$ decay rate. The updated value is \cite{PDG}
\beq\label{eq:f0}
|f_{\omega\pi}(0)|=(2.30 \pm 0.04) \gev^{-1}.
\eeq

As already mentioned, in the present paper we shall check the consistency of the data shown in Fig. \ref{fig:exp} by comparing the data in the decay region  with the allowed range  of the modulus imposed by unitarity and analyticity.  Mathematically, the problem amounts to deriving upper and lower bounds on $|f_{\omega\pi}(t)|$ for $t<t_+$, upon the class of functions $f_{\omega\pi}(t)$  analytic in the $t$-plane cut for $t\ge 4 m_\pi^2$, which satisfy the following conditions: ({\em i} ) their discontinuity is given by \eqref{eq:disc}  in the region  $t<t_+$, ({\em ii}) they satisfy the constraint \eqref{eq:L2}, and ({\em iii}) they satisfy the condition (\ref{eq:f0}).  The solution of this mathematical problem  will be given in the next section.

\section{Solution of the extremal problem}\label{sec:sol}
An  extremal problem of the type mentioned above was solved for the first time in   
\cite{Caprini:1981,  Caprini:1982nr} on the class of real analytic functions.  The generalization to functions which are not real analytic was investigated in detail in \cite{Anant:2014}. We do not repeat the whole proof here, but only outline the main steps and write down the solution.

The first step is to 
map the  $t$ plane cut along $t\ge t_+$  onto the unit disk $|z|\le 1$ in the  $z\equiv \tilde z(t)$ plane. We have adopted the conformal mapping
\beq\label{eq:z}
\tilde z(t)=\frac{1-\sqrt{1-t/t_+}}{1+\sqrt{1-t/t_+}},
\eeq 
which brings the origin of the $t$ plane to the origin of the $z$ plane,   $\tilde z(0)=0$. In the $z$-plane the elastic region $4 m_\pi^2\le t<t_+$ becomes the segment $x_\pi\le x<1$ of the real axis, where $x_\pi=\tilde z(4 m_\pi^2)>0$,  and the upper (lower) edges of the cut $t>t_+$ become the upper (lower) semicircles\footnote{Other mappings are obtained by changing the point that is mapped onto the origin of the $z$-plane. It can be shown \cite{Abbas:2010EPJA}  that the results  do not depend on the choice of the conformal mapping.}.

Further, we construct a so-called outer function \cite{Duren}, {\em i.e.} a function analytic and without zeros in $|z|<1$, its modulus  on $|z|=1$ being equal to $\sqrt{w(\tilde t(z)) |d\tilde t(z)/dz|}$, where $w(t)$ is the weight appearing in (\ref{eq:L2}) and $\tilde t(z)$ is the inverse of the function $\tilde z(t)$  defined in \eqref{eq:z}. The general expression of the outer functions is given in \cite{Duren} (see also the review \cite{Abbas:2010EPJA}). For weights $w(t)$ of the form  \eqref{eq:w}, we obtain for the outer function, denoted as $C(z)$, the exact analytic expression \cite{Abbas:2010EPJA}
\beq\label{eq:C}
C(z) =  (2\sqrt{t_+})^{1-c}  (1-z)^{1/2}  (1+z)^{c-3/2}.
\eeq  
From this expression it follows that $C(x)$ is real and positive for $-1< x<1$, which corresponds in the $t$-plane to the semiaxis $t<t_+$.

If we introduce now a new function $h(z)$ by
\beq\label{eq:h}
 h(z) = C(z)\,f_{\omega\pi}(\tilde t(z)),
\eeq 
 the condition \eqref{eq:L2} takes the simple form
\beq\label{eq:L2h}
 \D\frac{1}{2 \pi} \int_0^{2 \pi} d\theta |h(e^{i\theta})|^2 = I.
\eeq
 The function $h(z)$ is analytic in $|z|<1$ except for a  cut along the segment $(x_\pi, 1)$, where its discontinuity is
\beq\label{eq:disch}
\disc h(x)\equiv  \Delta (x)= C(x)\, \disc{f_{\omega\pi}(\tilde t(x))}.
\eeq
By expressing $h(z)$ as
\beq\label{eq:defg}
h(z)= \frac{1}{2\pi i}\int_{x_\pi}^1 \frac{\Delta (x)}{x-z} dx + g(z),
\eeq
the new function $g(z)$ is analytic in $|z|<1$, as its discontinuity across the cut vanishes:
\beq\label{eq:discg}
\disc{g(x)}=0, \quad \quad  x\in (x_\pi, 1).
\eeq
Since we consider in general form factors that are not real analytic, the function $g(z)$ is analytic, but its values on the real axis may be complex. 

We now express the available information on the form factor  as a number of constraints on the function  $g$.  By inserting (\ref{eq:defg}) in~\eqref{eq:L2h} we obtain the condition
\beq\label{eq:L2g}
\frac{1}{2 \pi}\int_0^{2 \pi} d\theta \left|\frac{1}{2\pi i}\int_{x_\pi}^1 \frac{\Delta (x)}{x- e^{i\theta}} dx + g(e^{i\theta})\right|^2 = I, 
\eeq
and using (\ref{eq:h}) and (\ref{eq:defg}) we write $g(0)$  as\footnote{Note that there is a misprint in the expression of $g(0)$ given in Eq. (24) of Ref. \cite{Anant:2014}.  It did not affect the results of \cite{Anant:2014} since the calculations were peformed with the correct expression.}
\beq\label{eq:g0}
g(0)= f_{\omega\pi}(0) C(0) -\frac{1}{2\pi i}\int_{x_\pi}^1 \frac{\Delta (x)}{x} dx. 
\eeq
The problem is to find 
the maximal allowed range of $|g(z_1)|$ at an arbitrary given point $z_1=\tilde z(t_1)$ in the interval $(x_\pi, 1)$, for functions $g(z)$ analytic in $|z|<1$ and subject both to the boundary condition~\eqref{eq:L2g} and the additional constraint~\eqref{eq:g0}.

It is useful to denote
\beq\label{eq:gz1}
g(z_1)= \xi,
\eeq
where $\xi$ is an unknown parameter. Then one can prove  (see for instance~\cite{Caprini:1982nr}) that the allowed  range of  $\xi$ is described by the inequality\footnote{This shows that the results remain the same if (\ref{eq:L2}) is replaced  by an inequality involving a quantity that majorizes  $I$.}
\beq\label{eq:range}
\mu_2^2(\xi) \leq I,
\eeq 
where $\mu_2^2(\xi)$ is the solution of the functional minimization problem
\beq\label{eq:mu2}
\mu_2^2(\xi) =\min_{g\in {\cal G}_\xi} \frac{1}{2 \pi}\int_0^{2 \pi} d\theta \left|\frac{1}{2\pi i}\int_{x_\pi}^1 \frac{\Delta (x)}{x- e^{i\theta}} dx + g(e^{i\theta})\right|^2,
\eeq
upon the class ${\cal G}_\xi$ of functions analytic in $|z|<1$, which satisfy the constraint~\eqref{eq:g0} and the additional condition~\eqref{eq:gz1} for a given $\xi$.

The constrained minimum norm problem \eqref{eq:mu2} was solved in \cite{Anant:2014}  by the technique of Lagrange multipliers, leading to a  solution written in compact form:
\begin{align}\label{eq:min}
\mu_2^2(\xi) &= \frac{1}{4 \pi^2} \int_{x_\pi}^1\int_{x_\pi}^1\frac{\Delta(x) \Delta^*(y)}{1-xy} dx\,dy + |g(0)|^2\nonumber\\
 &+ \frac{1-z_1^2}{z_1^2}\,|\xi -g(0)|^2.
\end{align}
By inserting (\ref{eq:min}) in (\ref{eq:range}) we obtain upper and lower bounds on the parameter $\xi$. Expressed in terms of the form factor $f_{\omega\pi}(t)$ by using Eqs. (\ref{eq:h}) and (\ref{eq:defg}), they lead to the inequalities \cite{Anant:2014}:
\begin{align}\label{eq:rangef}
 |f_{\omega\pi}(t)|&\leq \frac{\left|g(0)+\frac{1}{2\pi i}\int_{x_\pi}^1 \frac{\Delta (x)}{x-\tilde z(t)} dx\right|+\frac{\tilde z(t) I'}{\sqrt{1-\tilde z(t)^2}}}{C(\tilde z(t) )}\,,\nonumber\\
|f_{\omega\pi}(t)|&\geq \frac{\left|g(0)+\frac{1}{2\pi i}\int_{x_\pi}^1 \frac{\Delta (x)}{x-\tilde z(t)} dx\right|-\frac{\tilde z(t) I'}{\sqrt{1-\tilde z(t)^2}}}{C( \tilde z(t) )}\,,
\end{align}
where $\tilde z(t)\in (x_\pi, 1)$ is the image of the point $t$ in the $z$-plane,
and
\beq\label{eq:Iprime}
I'=\left[I-\frac{1}{4 \pi^2} \int_{x_\pi}^1\int_{x_\pi}^1\frac{\Delta(x) \Delta^*(y)}{1-xy} dx\,dy - |g(0)|^2\right]^{1/2}.
\eeq
We recall that $C(\tilde z(t))>0$ for $t<t_+$, which justifies its appearance outside the modulus sign in the denominator of (\ref{eq:rangef}). 

The upper and lower bounds (\ref{eq:rangef}) are calculable in terms of the input defined in the previous section. They determine an allowed interval for the modulus $|f_{\omega\pi}(t)|$ at every $t<t_+$.
From (\ref{eq:rangef}) it follows that, for a fixed weigth $w(t)$ in the $L^2$-norm constraint~\eqref{eq:L2}, the bounds  depend monotonically on the value of $I$: smaller values of  $I$ lead to narrower allowed intervals for  $|f_{\omega\pi}(t)|$ at $t<t_+$. We already took into account this property for a conservative estimate of $I$, as discussed in the previous section. 

It is useful to remark also that, since the last term in (\ref{eq:min}) is positive, from (\ref{eq:range}) and (\ref{eq:min}) we can write down the inequality 
\beq\label{eq:consist}
\frac{1}{4 \pi^2} \int_{x_\pi}^1\int_{x_\pi}^1\frac{\Delta(x) \Delta^*(y)}{1-xy} dx\,dy + |g(0)|^2\leq I,
\eeq
where $g(0)$ is defined in (\ref{eq:g0}) and $\Delta(x)$ is (\ref{eq:disch}). 
The inequality (\ref{eq:consist}) involves only input quantities and represents a necessary condition that must be satisfied by them. If it is violated,  the input is not consistent with analyticity and unitarity.

\section{Results}\label{sec:res}

We have investigated several suitable weights of the form (\ref{eq:w}) and checked that  they lead to similar results. The calculations reported below were done with the choice $c=2$,  which ensures a good suppression of the high energy part of the integral. 

As already mentioned, we have employed the discontinuity (\ref{eq:disc}) of the $\omega\pi$ form factor in the range $(4 m_\pi^2, t_+)$  from the recent dispersive treatment reported in \cite{Schneider:2012ez} and from the older work \cite{Koepp:1974da}. 
In \cite{Schneider:2012ez} the pion vector form factor $F_\pi(t)$ has been reconstructed from an Omn\`es representation \cite{Omnes} using as input the pion--pion phase shift $\delta_1^1(t)$ calculated from Roy equations in \cite{Madrid,CCL}. In \cite{Koepp:1974da}, the pion form factor was described by a Gounaris-Sakurai representation given in Eqs. (\ref{eq:GS})-(\ref{eq:Gamma}) of the Appendix. We have checked that the differences between the two representations of the pion form factor are very small and have a negligible influence on the results.

\begin{figure}[tbh]
\centering
\includegraphics[width=\linewidth, width=8.4cm]{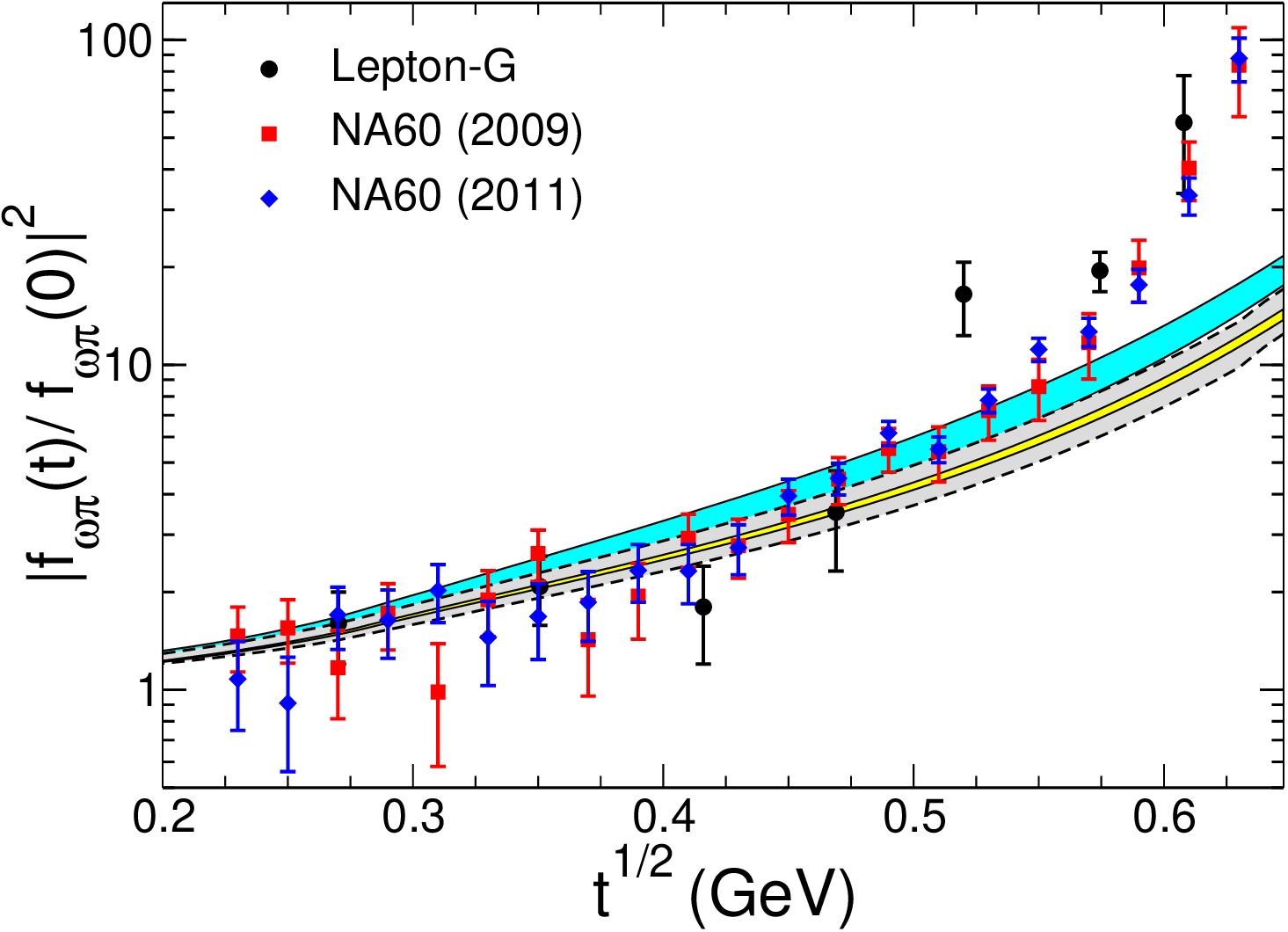}
\caption{Upper and lower bounds compared with experimental data on $|f_{\omega\pi}(t)/f_{\omega\pi}(0)|^2$.  Cyan band:  bounds calculated using in the discontinuity (\ref{eq:disc}) the partial wave amplitude $f_1(t)$ from \cite{Koepp:1974da}. Grey band: bounds calculated using in the discontinuity (\ref{eq:disc}) the amplitude  $f_1(t)$  from \cite{Schneider:2012ez}. The yellow band is the result of the dispersive calculation performed in \cite{Schneider:2012ez}.
The data are from Lepton-G~\cite{LeptonG}, NA60~(2009)~\cite{NA60} and NA60~(2011)~\cite{NA60new}.  
\label{fig:1} \vspace{0.7cm} }
\end{figure}

On the other hand, the differences in the partial wave $f_1(t)$ used in \cite{Koepp:1974da} and \cite{Schneider:2012ez} are sizable, and have a larger impact on the results.  In Fig. \ref{fig:1} we show the allowed bands determined by the upper and lower bounds  on the modulus squared (normalized to its value at $t=0$) in the part of the elastic region accessible experimentally in $\omega\to\pi^0\mu^+\mu^-$ decay, calculated using the expressions (\ref{eq:rangef}) with input from \cite{Koepp:1974da} and \cite{Schneider:2012ez}. The results  shown in Fig. \ref{fig:1} were obtained by varying the input value at $t=0$ inside the error bar given in (\ref{eq:f0}) and taking the weakest bounds, {\em i.e.} the largest allowed bands at each energy.  For comparison, we also show the result of the dispersive calculation performed in \cite{Schneider:2012ez}, and several experimental data from \cite{LeptonG,NA60,NA60new}. 

With the input discontinuity from \cite{Schneider:2012ez}, the allowed band is  consistent with the dispersion relation calculation performed in that work. The allowed band obtained with the partial wave $f_1(t)$ from \cite{Koepp:1974da} is shifted upwards and the two bands do not overlap.  For both inputs the upper bounds shown in Fig. \ref{fig:1} are significantly lower than the data from \cite{LeptonG,NA60, NA60new} in the region around 0.6 GeV. 

 We mention that the upper and lower bounds  shown in Fig. \ref{fig:1}  are much more stringent than the upper and lower bounds obtained in \cite{Anant:2014} with the same input on the discontinuity (\ref{eq:disc}), but with a model-independent condition on the modulus above the $\omega\pi$ threshold\footnote{For instance,   the allowed range of the ratio $|f_{\omega\pi}(t)/f_{\omega\pi}(0)|^2$ at 0.64 GeV obtained in \cite{Anant:2014} with input discontinuity from \cite{Schneider:2012ez} is  (1.6, 36.8), while the range predicted in this work with the same discontinuity is (11.3, 15.7).}.

\begin{figure}
\centering
\includegraphics[width=\linewidth, width=8.4cm ]{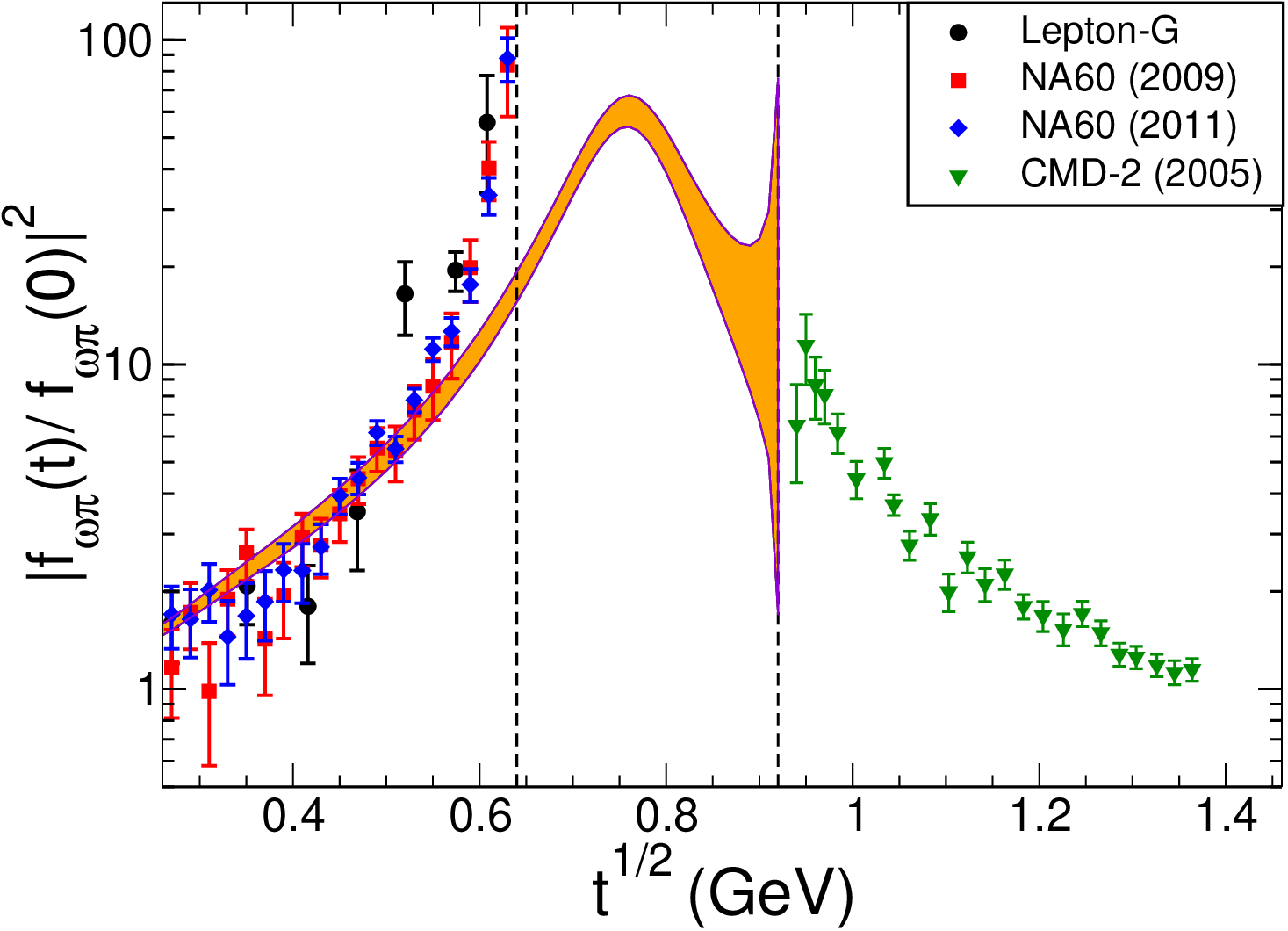}
\caption{Orange band: bounds on $|f_{\omega\pi}(t)/f_{\omega\pi}(0)|^2$ in the whole region $t<t_{+}$, obtained with the improved $N/D$ model. 
The data are from Lepton-G~\cite{LeptonG}, NA60~(2009)~\cite{NA60}, NA60~(2011)~\cite{NA60new} and CMD-2 (2005) \cite{CMD2-omegapi}. 
\label{fig:ext}\vspace{0.6cm} }

\end{figure}

 It is of interest to understand the origin of the difference between the two predictions shown in Fig. \ref{fig:1}. To this end, we have calculated the bounds using also the improved version of the $N/D$ model of \cite{Koepp:1974da}, which includes the effect of rescattering in the crossed channels as discussed in the Appendix. The improvement has the effect of shifting  the bounds downwards, towards  the band calculated with the input amplitude $f_1(t)$ from \cite{Niecknig:2012sj, Schneider:2012ez}, but the shift is small, of a few percents.  For illustration we present in Fig. \ref{fig:ext} the bounds calculated with the improved $N/D$ model for $f_1(t)$ in the whole range $t<t_+$.

\begin{figure}
\centering
\includegraphics[width=\linewidth, width=8.4cm]{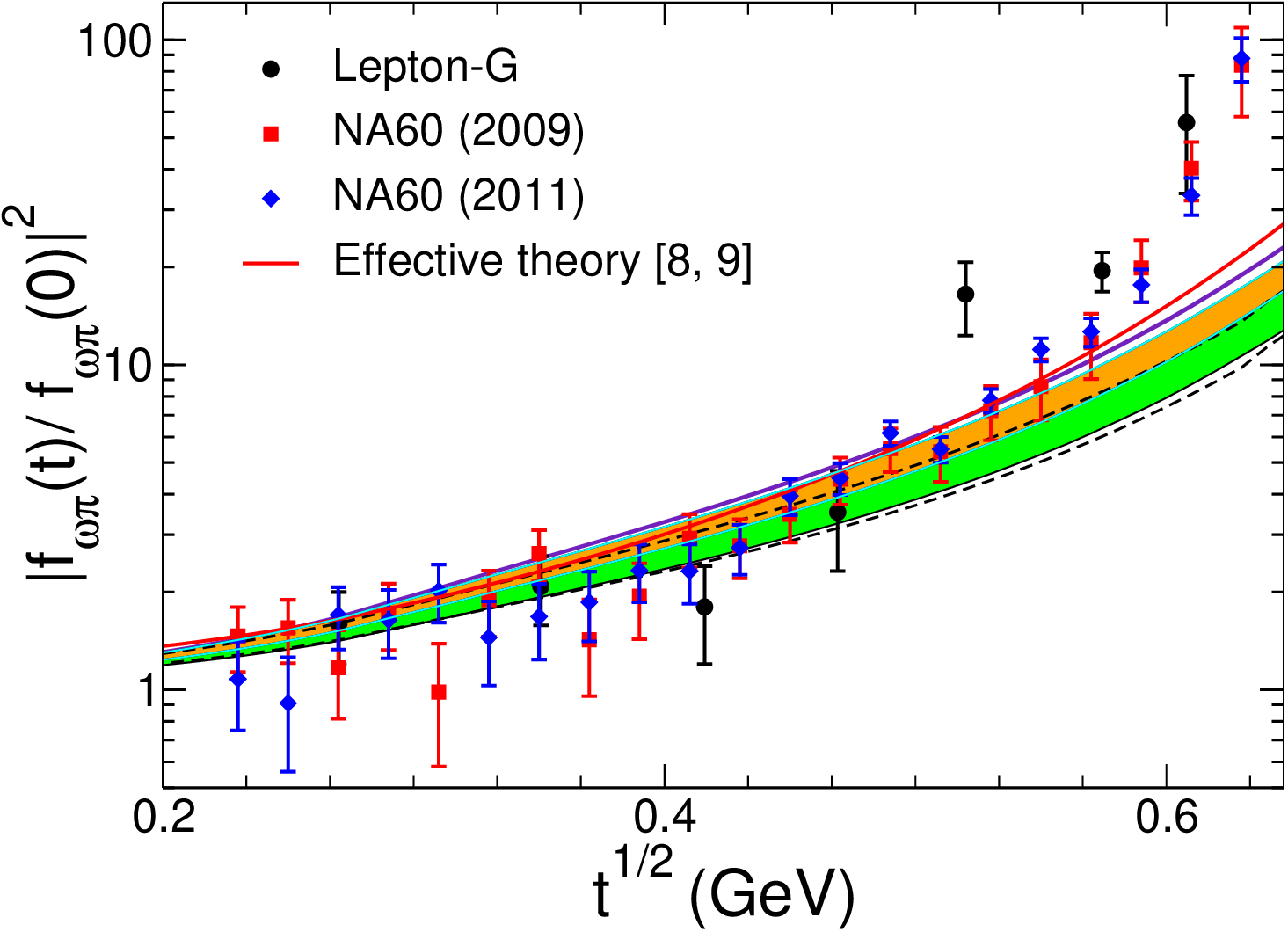}
\caption{ Upper and lower bounds on $|f_{\omega\pi}(t)/f_{\omega\pi}(0)|^2$ calculated using the improved $N/D$ model, with $g_{\omega\rho\pi}=11.5$ (green band)  $g_{\omega\rho\pi}=13.42$ (orange band) and  $g_{\omega\rho\pi}=15.5$ (indigo line). The dashed lines are the bounds calculated with the discontinuity (\ref{eq:disc})  from \cite{Schneider:2012ez}. The red line is the form factor calculated in \cite{Terschluesen:2010ik, Terschlusen:2012xw} with a low-energy effective theory.
\label{fig:3} \vspace{0.4cm} }
\end{figure}

As follows from Eqs. (\ref{eq:gdef}) and (\ref{eq:t1DR1}) of the Appendix, the $N/D$ model contains as input the dimensionless coupling constants $g_{\rho\pi\pi}$ and $g_{\omega\rho\pi}$. 
The  results presented in Figs. \ref{fig:1} and \ref{fig:ext} have been obtained using  the values from  \cite{Koepp:1974da}
\beq\label{eq:gconst}
g_{\rho\pi\pi}=5.96,\quad g_{\omega\rho\pi}=13.42.
\eeq
The quantity $g_{\rho\pi\pi}$ has not changed significantly over the last 40 years, the value (\ref{eq:gconst}) being fully consistent with the PDG-2014  width $\Gamma_\rho$, quoted below Eq. (\ref{eq:Gamma}). On the other hand, the coupling  $g_{\omega\rho\pi}$ is still not very well known. We note that the dimensionless constant $g_{\omega\rho\pi}$ used here is related to the similar  parameter used in \cite{Moussallam:2013, CMD2-omegapi}, which we denote as $\tilde g_{\omega\rho\pi}$ to avoid confusion,  by $g_{\omega\rho\pi}= m_\omega \tilde g_{\omega\rho\pi}$.  The values $\tilde g_{\omega\rho\pi}=(13.8 \pm 0.3)\,\gev^{-1}$ obtained in \cite{Moussallam:2013} from a global fit of  the $\omega\pi$ form factor, and $\tilde g_{\omega\rho\pi}=( 16.7 \pm 0.4\pm 0.6) \,\gev^{-1}$ derived in \cite{CMD2-omegapi} from a fit of the cross section of $e^+e^-\to\omega\pi^0$, correspond in our notation to the values $g_{\omega\rho\pi}=10.80 \pm 0.23$ and $g_{\omega\rho\pi}  =13.07 \pm 0.31 \pm 0.47$, respectively. We note that, while the above references consider parametrizations of  the $\omega\pi$ form factor, in the present analysis the coupling $g_{\omega\rho\pi}$ is an input parameter in the model of the amplitude $f_1(t)$. 

One might ask whether a suitable choice of the parameter $g_{\omega\rho\pi}$  can reduce the conflict between the calculated bounds and the experimental data.  We  have investigated this question using the improved $N/D$ model discussed in the Appendix. We remark first that for values $g_{\omega\rho\pi}>15.5$ the
inequality (\ref{eq:consist}), which expresses a consistency condition on the input,  is violated. Therefore, values of  $g_{\omega\rho\pi}$ larger than 15.5 are not  allowed, being inconsistent with the other input quantites and the general properties of analyticity and unitarity adopted here. 

In Fig. \ref{fig:3} we show the allowed bands calculated with three choices of the couplings $g_{\omega\rho\pi}$: the value (\ref{eq:gconst}), a smaller value,   $g_{\omega\rho\pi}=11.5$,  and the maximum allowed value $g_{\omega\rho\pi}=15.5$ mentioned above. In the latter case, when the the
inequality (\ref{eq:consist}) is saturated, the quantity $I'$ defined in (\ref{eq:Iprime}) is zero and  the upper and lower bounds written in (\ref{eq:rangef}) become equal. Therefore, as shown in Fig. \ref{fig:3}, the allowed band of the modulus shrinks in this case to a line.  The curves show that the bounds exhibit a monotonous dependence  on the value of $g_{\omega\rho\pi}$, the band obtained with $g_{\omega\rho\pi}=11.5$ being consistent with the allowed domain obtained using as input $f_1(t)$ from the calculation \cite{Niecknig:2012sj, Schneider:2012ez}. An important conclusion is that the disagreement with the experimental data around 0.6 GeV is preserved even if the coupling $g_{\omega\rho\pi}$ is increased up to its maximum allowed value.

We show also in Fig. \ref{fig:3} the $\omega\pi$ form factor calculated in \cite{Terschluesen:2010ik, Terschlusen:2012xw} within an effective field approach. It  exhibits a more rapid increase and slightly exceeds the highest allowed band shown in Fig. \ref{fig:3} near 0.6 GeV.

\section{Discussion and conclusions}\label{sec:disc} 
Our study has been motivated by the existence of certain discrepancies between the recent calculations  \cite{Schneider:2012ez, Danilkin:2014cra, Anant:2014} of the $\omega\pi$ electromagnetic transition form factor in the frame of dispersion theory, and the data measured
 from the $\omega\to\pi^0\gamma^*$ decay around 0.6 GeV. The present work differs from the previous calculations by the different input used above the $\omega\pi$ threshold $t_+$: while the investigations \cite{Schneider:2012ez, Danilkin:2014cra} are based on a standard dispersion relation requiring the knowledge of the discontinuity of the form factor along the whole cut, and the work \cite{Anant:2014} exploits a model-independent integral condition on the modulus, derived  from unitarity and perturbative QCD,   we have resorted to experimental data  obtained from $e^+e^-\to \omega\pi^0$. 
In order to reduce the bias due to the absence of data  at higher energies, we have implemented this information in a conservative way, as a weighted integral (\ref{eq:L2}) of the modulus squared.  

The aim of our study was to test the consistency of the experimental and theoretical information available on the $\omega\pi$ form factor in a parametrization-free approach.
 We have derived upper and lower bounds on the modulus  for $t<t_+$, using as input the discontinuity (\ref{eq:disc}) in its region of validity below the $\omega\pi$ threshold, and the condition  (\ref{eq:L2}) on the modulus above the  $\omega\pi$ threshold. Mathematically, the problem is of the type considered some time ago for real-analytic functions in \cite{Caprini:1981, Caprini:1982nr} and generalized recently to analytic functions which are not of real type in \cite{Anant:2014}. Since we used experimental data above $t_+$, the results obtained in the present work are much stronger than those obtained in \cite{Anant:2014}, where only a theoretical inequality on the modulus above $t_+$ was exploited.  

The main theoretical ingredient of the analysis is the partial wave $f_1(t)$ entering the discontinuity (\ref{eq:disc}). Therefore, it is of interest to establish the influence of various parameters entering this quantity on the final results. The amplitude  $f_1(t)$ calculated in \cite{Niecknig:2012sj,Danilkin:2014cra} is available only in numerical form and is not suitable for this purpose. We have considered therefore the older calculation based on $N/D$ formalism performed in \cite{Koepp:1974da}, which has the advantage of displaying in an explicit way the dependence on various parameters, and have improved it by a prescription suggested in \cite{Moussallam:2013} for including the 
effect of rescattering in the crossed channels.

Our study  has showed that including the rescattering  has the effect of shifting down the allowed band for the modulus of the form factor in the region $t<t_-$. However, in the frame of the $N/D$ model the effect is quite modest, of a few percents.  On the other hand, the  bounds are quite sensitive to the coupling  $g_{\omega\rho\pi}$, which enters as input in the calculation of  $f_1(t)$ in the $N/D$ formalism.  The  results obtained with $f_1(t)$ from \cite{Niecknig:2012sj, Schneider:2012ez} can be reproduced by using the value $g_{\omega\rho\pi}=11.5$ in the improved $N/D$ formalism. It turns out that values of $g_{\omega\rho\pi}$ larger than 15.5 are excluded, being inconsistent with the remaining elements of the input.  By increasing $g_{\omega\rho\pi}$, the allowed bands are pushed upwards. However, as shown in Fig. \ref{fig:3}, the narrow  band calculated with  the maximum allowed value of $g_{\omega\rho\pi}$ is still significantly lower than the experimental data from \cite{LeptonG,NA60,NA60new} near 0.6 GeV. 

Our  results reveal a clear conflict between the  experimental 
data on the modulus of the $\omega\pi$ form factor measured in the decay region $t<t_-$ from $\omega\to\pi^0\gamma^*$  and in the scattering region  $t>t_+$ from $e^+e^-\to\omega\pi^0$.  We note that possible discrepancies between the 
data on the modulus measured at energies below $t_-$ and above $t_+$ can be noticed also in the attempts to describe the form factor with specific
 parametrizations  \cite{Moussallam:2013}. In contrast,  no parametrization of the form factor was necessary in the present analysis.  
The present work confirms the conclusions of other recent dispersive analyses \cite{Schneider:2012ez, Anant:2014} and  brings further arguments in support of renewed experimental efforts to measure more precisely the $\omega$ conversion decays \cite{WASA, CLAS}.

\subsection*{Acknowledgments} 
I would like to thank B. Moussallam for very interesting discussions, and B. Ananthanarayan and B. Kubis for a pleasant collaboration on the work \cite{Anant:2014} and useful suggestions on the manuscript. This work was supported by  UEFISCDI under Contract Idei-PCE No 121/2011 and by the Ministry of Education under Contract PN No 09370102/2009.

\appendix*
\section{Improved $N/D$ treatment of the $\omega\pi \to\pi\pi$ amplitude }
The  $N/D$ model proposed in \cite{Koepp:1974da}  does not include the rescattering between all the final pions in the kinematical region where the $\omega$ decay to three pions is allowed. In this Appendix we briefly describe the model  and present a simple modification, which is able to capture  the characteristic features of the  full solution. 

For convenience, we use in this Appendix the notation of \cite{Koepp:1974da}.
The relation with the conventions used in the text is clear by comparing Eqs. (5.1) and (5.3)  of \cite{Koepp:1974da} with Eqs. (\ref{eq:disc}) and (\ref{eq:sigmaf}) of this paper, respectively.
The P partial wave amplitude of the scattering process (\ref{eq:scatt}), denoted  in \cite{Koepp:1974da} as $t^1(t)$, has dimensions of $\gev^{-2}$ and is related to the partial wave $f_1(t)$ of \cite{Niecknig:2012sj, Schneider:2012ez}  by
\beq\label{eq:f1t1}
t^1(t)=\frac{2}{3}\, m_\omega\, f_1(t).
\eeq
In the  $N/D$ formalism, the amplitude $t^1(t)$ is written as \cite{Koepp:1974da}
\beq\label{eq:tRL}
t^1(t)=t^1_L(t)+t^1_R(t),
\eeq
where $t^1_L(t)$ has only a left-hand cut and $t^1_R(t)$ has only a right hand cut for $t>4 m_\pi^2$.

The piece $t^1_L(t)$  was calculated in \cite{Koepp:1974da}  as 
\beq\label{eq:t1L}
t^1_L(t)= \frac{1}{2} \int_{-1}^1 dz \left[d_{00}^2(\theta)-d_{00}^0(\theta)\right] T_{su}(t, z),
\eeq
where $z\equiv \cos\theta$, $d_{m'm}^j(\theta)$ are elements of Wigner's $d$-matrix and
\beq\label{eq:Asu}
T_{su}(t, z)=-\frac{4}{3} \,g_{1} g_{2}  \left(\frac{1}{m_\rho^2-s(t, z )}+ \frac{1}{m_\rho^2-u(t,z)}\right)
\eeq
is the $\rho$-pole contribution in the crossed channels. The  dimensionless coupling constants are defined as \cite{Koepp:1974da}
\beq\label{eq:gdef}
g_1\equiv g_{\rho\pi\pi},\quad   g_2 \equiv g_{\omega\rho\pi}
\eeq 
and the Mandelstam variables  have the expressions
 \beq\label{eq:Mand}
 s(t,z)= R(t)+ K(t) z,\quad
 u(t,z)= R(t) - K(t)z,
\eeq
with
\beq\label{eq:RK}
R(t)=\frac{m_\omega^2 + 3 m_\pi^2-t}{2},\quad K(t)= 2 q(t)\,p(t), 
\eeq
where $q(t) $ and $p(t) $ are defined below Eqs. (\ref{eq:disc}) and (\ref{eq:sigmaf}), respectively.

The part $t_R^1(t)$  of the amplitude accounts for the rescattering in the direct channel. In the two-pion approximation, unitarity  gives 
\beq\label{eq:discR}
\disc [t^1 (t) \Omega^{-1}(t)]=0,\quad  t\ge 4 m_\pi^2,
\eeq
where  $\Omega(t)$  is the  Omn\`es function
\beq\label{eq:Omnes}
\Omega(t)= \exp\left[\frac{t}{\pi}\int_{4 m_\pi^2}^\infty \frac{\delta_1^1(t')}{t'(t'-t)}\,dt \right],
\eeq
which is analytic without zeros in the $t$-plane cut for $t>4 m_\pi^2$ and is normalized to $\Omega(0)=1$. It can be written above the cut as 
\beq\label{eq:Omnesmod}
\Omega(t+i\epsilon)= |\Omega(t)| e^{i \delta_1^1(t)},\quad\quad t\ge 4 m_\pi^2,
\eeq
where $\delta_1^1(t)$ is the phase shift  of the P wave of the elastic pion-pion amplitude. 

Since  $t^1_L(t)$ is regular for $t\ge 4 m_\pi^2$, from (\ref{eq:tRL}) and (\ref{eq:Omnesmod})  it follows that
\beq\label{eq:disctR}
\disc [t_R^1(t) \Omega^{-1}(t)]=- 2it^1_L(t) \Im\,[ \Omega^{-1}(t)].
\eeq
From the discontinuity one can reconstruct the function by means of a standard dispersion relation, written  in 
\cite{Koepp:1974da}  as  
\beq\label{eq:tR}
t_R^1(t) \Omega^{-1}(t)= a + \frac{t-t_0}{\pi} \int_{4 m_\pi^2}^\infty \frac{t^1_L(t') \sin \delta^1_1(t')} {|\Omega(t')|(t'-t_0)(t'-t)} dt,
\eeq
in terms of the unknown subtraction constant $a$.   Combined with (\ref{eq:tRL}), this  leads to 
\bea\label{eq:tOmnes}
t^1(t)&=&  \Omega(t) \left[\frac{t^1_L(t)}{\Omega(t)}+ a \right.\\ 
 &+ & \left. \frac{t-t_0}{\pi} \int_{4 m_\pi^2}^\infty \frac{t^1_L(t') \sin \delta^1_1(t')} {|\Omega(t')|(t'-t_0)(t'-t)} dt \right]\,.\nonumber
\eea

In \cite{Koepp:1974da}, instead of the Omn\`es function  $\Omega(t)$ a Gounaris-Sakurai  parametrization \cite{GS} was actually adopted,  which is a reasonable approximation on the right hand cut where it is employed. Thus, 
\beq\label{eq:GS}
\Omega(t)\Rightarrow GS(t) = \frac{D(0)}{D(t)},
\eeq
where $D(t)$ is written as  
\beq\label{eq:D}
D(t)= m_\rho^2-t-g(t) - i m_\rho \Gamma_\rho(t).
\eeq
In this relation
\beq\label{eq:Gamma}
\Gamma_\rho(t)=\frac{m_\rho}{\sqrt{t}} \left(\frac{q(t)}{q(m_\rho^2)}\right)^3\,\Gamma_\rho
\eeq
is the energy-dependent $\rho$ width defined in terms of the physical width $\Gamma_\rho=147.8 \pm 0.9\, \mev$  \cite{PDG},
and
\beq 
g(t)= \frac{m_\rho \Gamma_\rho}{q(m_\rho^2)}\left(
k(t)-k(m_\rho^2)-(t-m_\rho^2)k'(m_\rho^2)\right), \nonumber
\eeq
\beq\label{eq:kf}
k(t)=\frac{2 q(t)^3}{\pi \sqrt{t}} \ln\frac{2 q(t)+\sqrt{t}}{2 m_\pi}.
\eeq

Choosing the subtraction point at $t_0=m_\rho^2$, the behavior of $t^1(t)$  near $t= m_\rho^2$ implies
\beq\label{eq:a}
 a = \frac{4 }{3 }\, \frac{g_1 g_2}{D(0)}.
\eeq 
Then,  the representation (\ref{eq:tOmnes}) is written finally as \cite{Koepp:1974da}
\bea\label{eq:t1DR1}
t^1(t)&\!=\!&\frac{1}{D(t)} \left[t_L^1(t) D(t) + \frac{4}{3}\, g_{1} g_{2} \right. \\
&\!+\!& \left. \frac{m_\rho (t-m_\rho^2)}{\pi} \int_{4 m_\pi^2}^\infty \frac{t^1_L(t') \Gamma_\rho(t')} {(t'-m_\rho^2)(t'-t)} dt\right]. \nonumber
\eea

As follows from (\ref{eq:t1L}) and (\ref{eq:Asu}), $t_L^1(t)$ is real for $t>4 m_\pi^2$, which implies that the imaginary terms within the large parantheses compensate each other. Therefore, the phase of $t^1(t)$ is equal to the phase of the Omn\`es function $1/D(t)$, {\em i.e.} to the phase shift $\delta^1_1(t)$. In this model, $t^1(t)$ satisfies Watson theorem, and leads to a purely imaginary discontinuity (\ref{eq:disc}) of the $\omega\pi$  form factor. As shown in 
\cite{Niecknig:2012sj,Schneider:2012ez,Danilkin:2014cra},  these properties are no longer valid in the more rigorous treatments of the amplitude.

An obvious shortcoming of the model \cite{Koepp:1974da} is the fact that the amplitude $t^1_L(t)$ was calculated in terms of a $\rho$-meson exchange
neglecting the width of the $\rho$. In this approximation the $\omega$
meson is actually stable since its mass is lower that the mass of
$\rho\pi$ pair. To improve the model, a straightforward procedure would be to include a finite width for the $\rho$ poles in the denominators of (\ref{eq:Asu}).  We have adopted the  prescription proposed in \cite{Moussallam:2013}, where  finite-width resonance exchange amplitudes with correct  analyticity properties were obtained by replacing 
\beq\label{eq:presc}
\frac{1}{m_\rho^2-s(t,z) } \,  \Rightarrow  ~~\frac{1}{\pi}\int_{4 m_\pi^2}^\infty dx\, 
\frac{\sigma(x)}{x- s(t,z)},
\eeq 
and similarly for the $u$-channel contribution. The pole was replaced by a modified  Breit-Wigner expression which automatically ensures the absence of singularities in the complex plane except for a right-hand cut. As suggested in \cite{Moussallam:2013},  a reasonable choice for the spectral function
$\sigma(x)$  is the imaginary part of the Breit-Wigner propagator: 
\beq\label{eq:sigma}
\sigma(x)=\frac{m_\rho\Gamma_\rho(x)}{(m_\rho^2-x)^2 + m_\rho^2 (\Gamma_\rho(x))^2},
\eeq with $\Gamma_\rho(x)$ defined in (\ref{eq:Gamma}).
In the limit of zero width, $\Gamma_\rho\to 0$, when $\sigma(x)\to \pi \delta(x-m_\rho^2)$, the left side of (\ref{eq:presc}) is recovered.

By inserting the prescription (\ref{eq:presc}) in (\ref{eq:t1L})-(\ref{eq:Asu}), the integration upon $z\equiv \cos\theta$ can be performed exactly, leading to
\beq\label{eq:t1LBW}
t^1_L(t)= \frac{4 g_1 g_2}{3}\, \frac{1}{\pi}\int_{4 m_\pi^2}^\infty dx \,\sigma(x) 
\frac{F(Z)}{K(t)},
\eeq
where\beq\label{eq:FF}
F(y)=\frac{3}{4}\left[2 y +(1-y^2) \ln\frac{y+1}{y-1}\right],
\eeq
and
\beq
Z=\frac{x-R(t)}{K(t)},
\eeq
 with $R(t)$ and $K(t)$ defined in (\ref{eq:RK}).

The singularities of $t^1_L(t)$ in the complex plane arise from the singularities of the function $F(y)$ at $y=\pm 1$ produced by the logarithm, which  depend parametrically on $x$. When $x$ varies along the integration range in (\ref{eq:t1LBW}), the singularities describes paths in the complex $t$-plane and in principle can overlap with the $t$-channel unitarity cut along the real semiaxis $t\ge 4 m_\pi^2$.  The overlap can be avoided by a suitable prescription.  In the present study we have adopted the prescription proposed in \cite{BK}, which consists in  adding to  $m_\omega^2$ a small imaginary part, {\em i.e.} $m_\omega^2\to m_\omega^2+i \epsilon$, with $\epsilon>0$. With this prescription, we checked numerically that  the singularities of $t^1_L(t)$ do not cross the unitarity cut in the $t$-plane. Moreover,  the amplitude $t^1_L(t)$ has no discontinuity across the line $t\ge 4 m_\pi^2$, although it is no longer real on the unitarity cut. 

 The points $t=4  m_\pi^2$ and $t=t_\pm$, {\em i.e.} the physical thresholds and the pseudo-threshold $t_-$,  require special attention since there the function $K(t)$ defined in (\ref{eq:RK}) vanishes. By using the asymptotic expansion 
\beq\label{eq:asympt}
F(y)\sim \frac{1}{y} +\frac{1}{5 y^3}+\dots , \quad \quad |y| \gg 1,
\eeq
and the decrease $\sigma(x)\sim 1/x$, we have checked explicitly that in the present model $t^1_L(t)$ is regular at these points.

Since the amplitude $t^1_L(t)$ calculated from (\ref{eq:t1LBW}) has no discontinuity across the unitarity cut $t\ge 4 m_\pi^2$,  the representation (\ref{eq:t1DR1}) remains valid. However, as mentioned above,  $t^1_L(t)$ is complex for  $t\ge 4 m_\pi^2$. Therefore, from (\ref{eq:t1DR1}) it follows that the phase of the  amplitude $t^1(t)$ for $t\ge 4 m_\pi^2$ is no longer equal to the phase $\delta_1^1(t)$ of the function $1/D(t)$.   Watson theorem, which was valid  in the original $N/D$ model, is no longer valid now.  Moreover,  the amplitude is  not an analytic function of real type. These properties are satisfied of course by  the exact solution   $f_1(t)$  calculated  in 
\cite{Niecknig:2012sj,Schneider:2012ez,Danilkin:2014cra}.

It is useful to compare the simple improved $N/D$ model presented here with the exact amplitude calculated  by solving numerically  integral equations of the Khuri-Treiman type. An obvious feature of the  $N/D$ model  is the lack of symmetry between the direct ($t$) and the crossed ($s$ and $u$) channels. In fact, in the decay region the dynamics in the three two-pion channels must be the same.   In the Khuri-Treiman formalism, by iteratively solving the relevant integral equation,  the symmetry between the three channels is gradually increased.  This adjustment is not performed in the  $N/D$ approach, which has a rigid structure. However, by improving the description of the crossed channels in the frame of the $N/D$ model, the main features of the exact partial wave amplitude $f_1(t)$, namely the failure of Watson theorem and the breakdown of the reality property,  appear in a natural way.

\end{document}